\documentclass[aps,prb,showpacs,twocolumn,superscriptaddress]{revtex4}
\usepackage{bm,color,amsmath,amssymb,mathrsfs,latexsym,graphicx,psfrag}

\newcommand{\braket}[2]{\left\langle #1 | #2 \right\rangle}
\newcommand{\bra}[1]{\left\langle#1\right|}
\newcommand{\ket}[1]{\left|#1\right\rangle}

\newcommand{\of}[1]{\!\left(#1\right)}
\newcommand{\sqof}[1]{\left[#1\right]}
\newcommand{\cuof}[1]{\left\{#1\right\}}

\newcommand{\up}{\uparrow}
\newcommand{\down}{\downarrow}

\newcommand{\abs}[1]{\left|#1\right|}

\newcommand{\expect}[1]{\left\langle#1\right\rangle}

 \newcommand{\Zcom}{\mathcal{Z}}

\newcommand{\Omegavec}{\bm{\Omega}}
\newcommand{\Uvec}{\bm{U}}
\newcommand{\Vvec}{\bm{V}}
\newcommand{\Wvec}{\bm{W}}

\newcommand{\Svec}{\bm{S}}

\newcommand{\cg}[6]{C\!\,_{#1}^{#2} \!\,_{#3}^{#4} \!\,_{#5}^{#6}}

\def\bd{\begin{displaymath}}
\def\ed{\end{displaymath}}
\def\be{\begin{equation}}
\def\ee{\end{equation}}
\def\bea{\begin{eqnarray}}
\def\eea{\end{eqnarray}}
\def\bi{\begin{itemize}}
\def\ei{\end{itemize}}
\def\bn{\begin{enumerate}}
\def\en{\end{enumerate}}

\def\ie{{\it i.e.},\ }
\def\eg{{\it e.g.}\ }

\newcommand{\vecbf}[1]{\mathbf{#1}}

\begin{document}
\title{Parent Hamiltonian for the Chiral Spin Liquid}

\author{Ronny Thomale}
\affiliation{Institut f\"ur Theorie der Kondensierten Materie, 
Universit\"at Karlsruhe, 76128 Karlsruhe, Germany}
\author{Eliot Kapit}
\affiliation{Department of Physics, Cornell University, Ithaca, NY 14853 }
\author{Darrell F. Schroeter}
\affiliation{Department of Physics, Reed College, Portland, OR 97202}
\author{Martin Greiter}
\affiliation{Institut f\"ur Theorie der Kondensierten Materie, 
Universit\"at Karlsruhe, 76128 Karlsruhe, Germany}

\pagestyle{plain}

\date{\today}

\begin{abstract}
%
  We present a method for constructing parent Hamiltonians for the
  chiral spin liquid.  We find two distinct Hamiltonians for which the
  chiral spin liquid on a square lattice is an exact zero-energy
  ground state.  We diagonalize both Hamiltonians numerically for
  16-site lattices, and find that the chiral spin liquid, modulo its
  two-fold topological degeneracy, is indeed the unique ground state
  for one Hamiltonian, while it is not unique for the other.

\end{abstract}

\pacs{75.10.Jm, 05.30.Pr, 71.10.Hf}

\maketitle

\section{Introduction}
\label{sec:int}

The first notion of fractional excitations in condensed matter physics
goes back to the appearance of soliton mid-gap modes in
polyacetylene~\cite{su-79prl1698}, where the effective net charge of
one kink excitation is $e/2$, \ie one half of the electron charge. At
a similar time, the field of fractional statistics, founded in the
work by Leinaas and Myrheim~\cite{leinaas-77ncb1}, attained broad
attention due to the work by Wilczek in
1982~\cite{wilczek82prl1144,wilczek82prl957}. In strongly-correlated
many-body systems, the phenomenon of fractionalization, where the
elementary excitations of the system carry only a fraction of the
quantum numbers of the constituents, has become known to occur in a
variety of cases.

The first physical system in which fractional excitations and the
associated fractional statistics have been discussed on a unified
footing is the fractional quantum Hall
effect~\cite{laughlin83prl1395,halperin84prl1583, arovas-84prl722}
(FQHE). 
There, the quantum statistics of the anyonic quasiparticles can be
understood in terms of a generalized Berry's phase~\cite{Wilczek90},
which is acquired by the wave function as 
quasiparticles wind around each other.  This is a sensible concept in
two dimensions only where one can uniquely define a winding number for
the braiding.  In the FQHE, the fractional statistics is
known to occur in the presence of a magnetic field violating parity
(P) and time-reversal (T) symmetry. In recent years, there have been
tremendous efforts to study the fractional excitations of the FQHE
experimentally in order to confirm the prediction from theory and
to validate fractional statistics as a concept being realized in
nature. This, however, has remained inconclusive in certain aspects
and thus is still a subject of current discussion and
work~\cite{goldman-95s1010,depicciotto-97n162,reznikov-99n238,camino-05prl246802,feldman-07prb085333}.

Later, the concept of fractional statistics has been found to occur in
one-dimensional spin-1/2 antiferromagnets, where it can be defined in
terms of a generalized Pauli principle obeyed by the
excitations~\cite{haldane91prl937} and, as shown recently, by a phase
the wave function acquires when two spinons move through each
other~\cite{greiter09prb064409}.  The fractional charge of the
quasiparticles in the FQHE corresponds to the spin $1/2$ of the
elementary spinon excitations in these systems, which is fractional as
the Hilbert space is built up by spin flips which carry spin $1$.  As
one-dimensional systems are amenable to a host of exact methods, many
exactly solvable models exhibiting this behavior
exist~\cite{haldane88prl635,shastry88prl639,haldane91prl1529,haldane-92prl2021}.
In particular, various properties of fractional excitations in spin
chains have been observed
experimentally~\cite{tennant-93prl4003,coldea-97prl151,kim-06np397}.

In general, it appears to be that $\mathrm{P}$ and $\mathrm{T}$
violation is intimately related to the occurrence of excitations
obeying fractional statistics in two dimensions, which both applies to
quasiparticles in the FQHE and spinons in a quantum antiferromagnet.
These symmetries may be explicitly broken as in the FQHE or generated
by spontaneous symmetry breaking.  For two-dimensional
antiferromagnets, the concept of fractional excitations is less
established than for the one-dimensional
case~\cite{nayak-01prb064422}. In particular, finding solvable
theoretical models in which the phenomenon occurs has been one
predominant area of research in the field.  Significant progress has
been accomplished for dimer models~\cite{misguich-02prl137202,
  moessner-01prl1881}.

In addition to important questions with regard to the general
principle underlying fractional statistics, two-dimensional spin
liquids are of special interest with regard to investigation of the
hypothesized link between fractionalization and high-$T_{\text{c}}$
superconductivity~\cite{anderson87s1197,kivelson-87prb8865}.
Moreover, in many systems where fractionalization occurs, there is the
ambition to use the topological degeneracy contained in these systems
for quantum computing, where topological information can serve as a
quantum bit with negligibly small local decoherence
rates~\cite{Kitaev02ap2}.

The paradigmatic state for a $S=1/2$ spin liquid is the chiral spin
liquid (CSL) introduced by Kalmeyer and
Laughlin~\cite{kalmeyer-87prl2095,kalmeyer-89prb11879}, which is
constructed to spontaneously violate the symmetries $\mathrm{P}$ and
$\mathrm{T}$, and can be defined on any regular lattice including both
bipartite and non-bipartite lattices.  The universality class of
chiral spin liquid states, and in particular the order parameter and
the topological degeneracy~\cite{wen-89prb7387}, was defined by Wen,
Wilczek, and Zee~\cite{wen-89prb11413}.  A CSL state has also been
constructed by Yao and Kivelson~\cite{yao-07prl247203} in the Kitaev
model~\cite{Kitaev06ap2} on a Fisher lattice, \ie a honeycomb lattice
of triangles.  Recently, a family of non-Abelian CSL
states~\cite{greiter-09prl207203} has been proposed for general spin
$S$, whose wave functions correspond to the bosonic Read-Rezayi series
of FQH states~\cite{Read-99prb8084}.  The non-Abelian statistics of
the spinons has even been conjectured to be a general property of spin
$S$ antiferromagnets~\cite{greiter-09prl207203}.


As in the one-dimensional case mentioned above, the spinons in the CSL
exhibit quantum-number fractionalization and carry only half the spin
of the bosonic spin excitations in conventional magnetically-ordered
systems, which carry spin~$1$. Whereas the spinon appears to be the
fundamental field describing excitations in two-dimensional $S=1/2$
antiferromagnets in general, an effective description by magnon-like
excitations proved rather adequate for the generic model. The reason
for this is that the confinement between the spinons is generically so
strong that the underlying excitation structure is mostly suppressed.
In the CSL, however, the spinons are deconfined.  The model is hence
ideally suited to study fractional quantization of spinons in two-dimensional
antiferromagnets.  In
spite of its promising properties, for nearly two decades since its
emergence, the CSL lacked a microscopic model where it is realized.



In this article, we develop 
an analytical method for the construction of parent Hamiltonians for
the CSL.  The method relies explicitly on the singlet property of the
CSL, as this allows for a spherical tensor decomposition of the
destruction operator we introduce. From different tensor components,
we construct two different parent Hamiltonians, which annihilate the
CSL and hence have it as a zero-energy ground state.  One of the
Hamiltonians has been presented in a Letter
previously~\cite{schroeter-07prl097202}; both Hamiltonians contain
6-body interactions.  One of the key issues we address here is whether
the CSL is the {\it only} ground state of these Hamiltonians.  To
answer this question, we perform exact diagonalization studies of both
models for a 16-site square lattice.  In particular, we introduce an
adapted Kernel sweeping method, which allows for an efficient
numerical implementation of the complex and technically cumbersome
Hamiltonians we investigate.  We find that the model we introduced
previously has indeed the CSL as its (modulo the two-fold topological
degeneracy) unique ground state.  For the other Hamiltonian we
present, however, we find that the CSL is not the unique ground state.
Hence only the former model is useful for further analysis of \eg the
spinon spectrum.

The paper is organized as follows. In Section~\ref{sec:groundstate},
we review the chiral spin liquid ground state and its basic
properties.  After outlining the general construction scheme for the
Hamiltonians in Section~\ref{sec:gencon}, we formulate a destruction
operator for the CSL state in Section~\ref{sec:destruction} and
exploit the spin rotational invariance of the CSL state to decompose
the destruction operator into its spherical tensor components, which
annihilate the CSL state individually.  The proof that the destruction
operator annihilates the CSL ground state is given in
Section~\ref{sec:proof}.  In Section~\ref{sec:kernelsweeping}, we
introduce a Kernel sweeping method to compute the CSL Hamiltonians.
We present the method in detail and emphasize its applicability to
efficiently compute $n$-body interactions for finite-size exact
diagonalization studies.  The numerical results obtained with this
method are discussed in Section~\ref{sec:numerics}.  We conclude this 
work with a summary in Section~\ref{sec:conclusion}.

\section{Chiral Spin Liquid}
\label{sec:groundstate}

\begin{figure}[t]
\centerline{\includegraphics[scale=0.33]{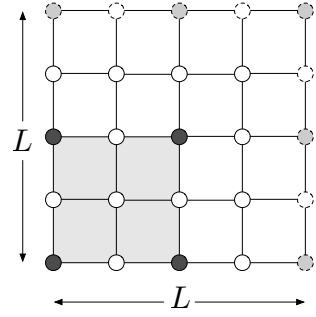}}
\caption{The model is defined on a square lattice length $L$ on a side
  such that the total number of sites is given by $N = L^2$.  The
  image shows the lattice for $N = 16$. The shaded circles (including
  the origin) indicate those lattice sites for which $G\of{z} = -1$
  and the open circles those sites for which $G\of{z} = +1$. The sites
  on which $G\of{z} = -1$ define a sublattice with twice the original
  lattice spacing; the doubled unit-cell is shown as the shaded region
  in the figure surrounding the origin. }
\label{fig:lattice2}
\end{figure}

 The CSL was originally conceived by
D.H.~Lee as a spin liquid constructed by condensing the bosonic spin
flip operators on a  two-dimensional lattice into a FQH liquid at Landau level
filling factor $\nu =1/2$.  The ground state wave function for a
circular droplet with open boundary conditions, on a square lattice
with lattice constant of length one, is given
by~\cite{kalmeyer-87prl2095, kalmeyer-89prb11879}
\begin{equation}
  \label{psiplane}
  \braket{z_1 \cdots z_{M}}{\psi}=
  \prod_{j<k}^M\,(z_j-z_k)^2\;\prod_{j=1}^M\,G(z_j)\,e^{-\frac{\pi}{2}|z_j|^2} \, ,
\end{equation}
where $M = N/2$.  The $z$'s in the above expression are the complex
positions of the up-spins on the lattice: $z=x+iy$, with $x$ and $y$
integer.  $G(z)=(-1)^{(x+1)(y+1)}$ is a gauge factor, which ensures
that \eqref{psiplane} is a spin singlet (see Fig.~\ref{fig:lattice2}).
Lattice sites not occupied by $z$'s correspond to down-spins.

For our purposes, it is propitious to choose periodic boundary
conditions (PBCs) with equal periods $L_1=L_2=L$, $L$ even, and with
$N=L^2$ sites.  Following Haldane and Rezayi\cite{haldane-85prb2529},
the wave function for the CSL then takes the form
\begin{eqnarray}
    \braket{z_1 \cdots z_{M}}{\psi}\! &\!=\!&\!
    \prod_{\nu =1}^2 \vartheta_1\of{\frac{\pi}{L}\sqof{\Zcom-Z_\nu}}
    \prod_{j<k}^M \vartheta_1\of{\frac{\pi}{L}\sqof{z_j-z_k}}^2 \nonumber\\
   &\cdot\!&\! \prod_{j=1}^M G(z_j)\,e^{\frac{\pi}{2}(z_j^2-|z_j|^2)},
  \label{eq:wavefunction}
\end{eqnarray}
where $\vartheta_1(w)=-\vartheta_1(-w)
\equiv\vartheta_1\of{w|e^{-\pi}}$ is the odd Jacobi theta
function~\cite{AbramowitzStegun65}.  The zeros for the center-of-mass
coordinate $\Zcom=\sum_j z_j$ must lie in the principal region
$0\leq\text{Re}\of{Z_1}<L$, $0\leq\text{Im}\of{Z_1}<L$ and satisfy
$Z_1+Z_2=L+iL$; the freedom to choose $Z_1$ reflects the topological
degeneracy and yields two linearly-independent ground states for the
CSL.  These states are spin singlets, are invariant under lattice
translations, and are strictly periodic with regard to the PBCs.

\section{General Method}
\label{sec:gencon}
In order to construct a parent Hamiltonian for the chiral spin liquid,
one first derives the destruction operators for the ground state.  In
our formulation, the destruction operators are constructed from a set
of operators $\omega_j$ where $j = 1, \ldots, N$ indexes the
lattice sites.  The operators $\omega_j$, to be introduced in
Section~\ref{sec:destruction} below, are not themselves destruction
operators, but have the property that, acting on the ground state, they
produce a result independent of the site index $j$: $\omega_i \,
\ket{\psi} = \omega_j \, \ket{\psi}$.  Therefore, once the above
result is established in Section~\ref{sec:proof}, it follows that the
difference of any two of the operators is a destruction operator for
the ground state: $d_{ij} = \omega_i - \omega_j$.

In order to construct a sensible parent Hamiltonian, one must
minimally demand that it be a translationally-invariant scalar
operator.  In order to put the Hamiltonian in this form, it is shown
in Appendix~\ref{app:decomp} that the operators may be written as
$\omega_j = \Omega_j^0 + \mho_j^0$ where $\Omegavec_j$ and
$\bm{\mho}_j$ are vector and third-rank spherical tensor operators
respectively and where the $0$ superscript indicates the component in
spherical notation.  The operators $\Omegavec_j$ and $\bm{\mho}_j$ are
given explicitly in terms of spin operators in
Sections~\ref{sec:vectordestroy} and \ref{sec:tensordestroy}.

As is discussed in detail in Section~\ref{sec:destruction}, the
Wigner-Eckhart theorem guarantees that all components of the operators
$\vecbf{D}_{ij} = \Omegavec_i - \Omegavec_j$ as well as
$\bm{\mathcal{D}}_{i j} = \bm{\mho}_i - \bm{\mho}_j$ are destruction
operators for the chiral spin liquid ground state so long as the
reducible tensor operator $d_{ij}$ is.  One can then construct
Hamiltonians based on either set of operators: \bea H =
\sum_{\expect{i \, j}} \vecbf{D}_{i j}^{\dagger} \cdot \vecbf{D}_{i j}
	\label{eq:hvec}
\eea
for the vector destruction operators or 
\bea
	H = \sum_{\expect{i \, j}} \sum_{\nu = -3}^{3}  \of{\mathcal{D}_{i j}^{\nu}}^{\dagger} \,  \mathcal{D}_{i j}^{\nu} 
	\label{eq:htense}
        \eea for the rank-$3$ spherical tensor operators.  Either
        Hamiltonian is a scalar and is translationally invariant, both
        of these properties guaranteed by the construction.
        Additionally, since the Hamiltonians are positive
        semi-definite, the chiral spin liquid is a ground state of the
        model.  It should be noted that these models are not
        themselves unique as one could include any coefficients
        $J_{ij}$ into the sums of Eqs.~\ref{eq:hvec} and
        \ref{eq:htense} and remove the restriction that only
        nearest-neighbor sites are summed over.  These two models do,
        however, represent the simplest models from each class.

        In Section~\ref{sec:kernelsweeping}, a numerical method is
        developed for performing the exact diagonalization of these
        Hamiltonians that can handle the large number of interactions
        efficiently.  This method is used in
        Section~\ref{sec:numerics} to show that the model given by
        Eq.~\ref{eq:hvec} has exactly two ground states, as
        expected due to the topological degeneracy of the chiral spin
        liquid on a torus, and that these states are precisely the
        chiral spin liquid ground states given in
        Section~\ref{sec:groundstate} above.  Adopting the same
        procedure, the Hamiltonian given in Eq.~\ref{eq:htense}
        is shown to have a larger ground-state manifold which is not
        exhausted by the chiral spin liquid ground states.

\section{Annihilation operator for the Chiral Spin Liquid}
\label{sec:destruction}

The Hamiltonian which stabilizes the chiral spin liquid is generated
by first finding a set of operators $\omega_i$, where $i$ is a site
index.  These operators are not themselves destruction operators, but
the bond operators $\omega_i - \omega_j$, where $i$ and $j$ are any
two distinct sites, will be shown to destroy the CSL ground state.
The operators may be written as $\omega_j = \omega_j^+ - \omega_j^-$
where $\omega_j^+ = T_j + V_j$ and \bea T_j & = & \frac{1}{2} \sum_{i
  \, k \neq j}' K_{i j k} \, S_j^+ \, S_k^- \, \of{\frac{1}{2} +
  S_i^z}
	\label{eq:tjdef} \\
	V_j & = & \sum_{i \neq j} U_{i j} \, \of{\frac{1}{2} + S_i^z} \, \of{\frac{1}{2} + S_j^z} \, .
	\label{eq:vjdef}
        \eea The two sets of coefficients $U_{i j}$ and $K_{i j k}$
        are defined in Section~\ref{sec:coeff} below and the prime on
        the sum indicates that one must exclude the coincidences of
        $i$ and $k$.

        The operator $\omega_j^-$ is related to $\omega_j^+$ by a
        $\pi/2$ rotation about the $x$-axis that maps $S_z$ and $S_y$
        into $-S_z$ and $-S_y$.  This means that the entire operator
        $\omega_j$ is given by
\begin{widetext}
\bea
	\omega_j = \sum_{i \, k \neq j}' K_{i j k} \, \sqof{\frac{1}{2 \, i} \, \of{\Svec_j \times \Svec_k}^z + \of{\Svec_j \cdot \Svec_k} \, S_i^z - S_i^z \, S_j^z \, S_k^z} + \sum_{i \neq j} U_{i j} \, S_i^z \, .
	\label{eq:omegaexplicit}
\eea
\end{widetext}
In writing down Eq.~\ref{eq:omegaexplicit}, the fact that
$\sum_{i \neq j} U_{i j} = 0$, has been employed.  This will be
demonstrated in Section~\ref{sec:coeff} below.  While the operators
$\omega_i$ are not themselves destruction operators for the CSL ground
state, it will be shown in Section~\ref{sec:proof} that $d_{i j} =
\omega_i - \omega_j$ is a destruction operator for the ground state
for any choice of $i$ and $j$.

The operators $\omega_j$ are reducible and can be decomposed into
irreducible tensor operators, in this case of ranks $1$ and $3$.  From
Eq.~\ref{eq:omegaexplicit} it is clear that every term except for
the $S_i^z \, S_j^z \, S_k^z$ term is the 0 (or $z$) component of a
rank-1 (vector) operator.  This final term can be decomposed into
rank-3 and vector components.

It is straightforward to show that if an operator $d$ is a destruction
operator for the CSL ground state, then each of its irreducible
components are as well.  This is because the Wigner-Eckhart theorem
tells us that acting with an operator $T^j_m$ on a state $\ket{n \, q
  \, m_q}$ with angular momentum $q$ and $z$-component $m_q$ gives
\bea T^j_m \, \ket{n \, q \, m_q} = \sum_{j' \, m'}
\cg{j}{m}{q}{m_q}{j'}{m'} \, \ket{n' \, j' \, m'} \, , \eea where $n$
and $n'$ are any quantum numbers other than angular momentum.  Since
the CSL is a spin singlet: $q = m_q = 0$, it follows that there is
only a single non-zero term in the above sum corresponding to $j' = j$
and $m' = m$.  This means that by decomposing the destruction operator
for the ground state $d$ into its tensor components, which may be
written $d = \sum_j \, a_j \, T^j_0$, acting on the ground state to
obtain \bea 0 = d \, \ket{\psi} = \sum_{j} a_j \, \ket{n' \, j \, 0}
\, , \eea and noting that states with different values of $j$ are
necessarily orthogonal, it immediately follows that each of the states
in the sum are themselves zero and hence the operators $T^j$ are
destruction operators for the ground state.  In
Sections~\ref{sec:vectordestroy} and \ref{sec:tensordestroy} we give
two classes of operators that are obtained from the reducible tensor
operator $\omega_j$ in Eq.~\ref{eq:omegaexplicit}.

\subsection{Vector destruction operator}
\label{sec:vectordestroy}

As shown in Appendix~\ref{app:decomp}, the operator $S_i^z \, S_j^z \,
S_k^z$ may be written as the sum of the $0$-components of a vector and
a third-rank tensor.  The vector component is given by \bea
\frac{1}{5} \, \sqof{\of{\Svec_i \cdot \Svec_j} \, \Svec_k^z +
  \of{\Svec_j \cdot \Svec_k} \, \Svec_i^z + \of{\Svec_k \cdot \Svec_i}
  \, \Svec_j^z}
	\label{eq:vecbit}
        \eea and, working from Eq.~\ref{eq:omegaexplicit}, the
        vector operator $\Omegavec_j$ is given by
\begin{widetext}
  \begin{eqnarray} 
    \Omegavec_j = \sum_{i, k \neq j}' K_{ijk} \, \sqof{ \frac{1}{2
        \, i} \, \of{\Svec_j \times \Svec_k} + \frac{4}{5} \,
      \of{\Svec_j \cdot \Svec_k} \, \Svec_i - \frac{1}{5} \, \of{\Svec_k
        \cdot \Svec_i} \, \Svec_j - \frac{1}{5} \, \of{\Svec_i \cdot
        \Svec_j} \, \Svec_k} + \sum_{i \neq j} U_{ij} \, \Svec_i \, .
	\label{eq:omegavecdef}
\end{eqnarray}
\end{widetext}
Since $\Omegavec_i - \Omegavec_j$ is a destruction operator for the
ground state, it immediately follows that one may construct a
Hamiltonian for which the chiral spin liquid is the exact ground state
as \bea H = \sum_{\expect{i \, j}} \, \of{\Omegavec_i -
  \Omegavec_j}^{\dagger} \cdot \of{\Omegavec_i - \Omegavec_j} \, ,
	\label{eq:hamvec}
        \eea where the sum runs over all nearest-neighbors on the
        lattice.  By construction, the Hamiltonian is a scalar
        operator and translationally invariant.

        However, note that there is nothing restricting possible
        models to run only over next-nearest neighbors.  Rather, one
        can consider any combination of bond-operators (including
        arbitrary coefficients so long as one maintains positive
        semi-definiteness in $H$) in constructing a parent Hamiltonian
        for the CSL.

\subsection{Tensor destruction operator}
\label{sec:tensordestroy}

It is also possible to create a set of third-rank tensor destruction
operators.  As shown in Appendix~\ref{app:decomp}, the operator $S_i^z
\, S_j^z \, S_k^z$ may be fully decomposed into the $0$-components of
a vector operator (given in Eq.~\ref{eq:vecbit}) and a third-rank
tensor operator, which is necessarily just the difference between
$S_i^z \, S_j^z \, S_k^z$ and the operator in
Eq.~\ref{eq:vecbit}.  This gives a destruction operator whose
$0$-component is
\begin{widetext}
\bea
	\mho_j^0 = - \frac{1}{\sqrt{10}} \sum_{i, k \neq j}' K_{i j k} \, \sqof{{\, \of{\Svec_i \cdot \Svec_j} \, S_k^z + \of{\Svec_j \cdot \Svec_k} \, S_i^z + \of{\Svec_k \cdot \Svec_i} \, S_j^z} - 5 \, S_i^z \, S_j^z \, S_k^z} \, .
	\label{eq:mhozero}
\eea
\end{widetext}
The other components are straightforward to obtain (see
Appendix~\ref{app:decomp}) and one may again use these operators to
form a Hamiltonian for the chiral spin liquid according to \bea H =
\sum_{\expect{i \, j}} \sum_{\nu = -3}^3 \, \of{\mho^{\nu}_i -
  \mho^{\nu}_j}^{\dagger} \of{\mho^{\nu}_i - \mho^{\nu}_j} \, .
	\label{eq:hamtense}
        \eea The Hamiltonian in Eq.~\ref{eq:hamtense} has two
        significant advantages over the model in Eq.~\ref{eq:hamvec}:
        it depends only on one set of coefficients ($K_{i j k}$ but
        not $U_{i j}$) and, because the operator in the sum in
        Eq.~\ref{eq:mhozero} is symmetric under interchange of $i$ and
        $k$, one may replace $K_{i j k}$ by $A_{i j k} = \of{K_{i j k}
          + K_{k j i}} / 2$ where the new coefficients are manifestly
        symmetric in interchange of the first and third indices.  
       Unfortunately, it turns out that the CSL is not the only ground state of this model, as will 
     be discussed in detail in Section~\ref{sec:numerics}.

\subsection{Coefficients}
\label{sec:coeff}

The coefficients appearing in Eq.~\ref{eq:omegaexplicit} are
functions of the distance between the sites of the form $K_{i j k} =
K\of{z_k - z_j, z_i - z_j}$ where \bea K\of{x,y} = \frac{1}{N/2 - 1}
\, \lim_{R \rightarrow \infty} \sum_{0 \leq z_0 \leq R} \frac{P\of{x -
    z_0, y}}{x - z_0},
	\label{eq:Kdef}
        \eea and the sum over $z_0$ is a sum over all lattice
        translations: $z_0 = \of{m + i \, n} \, L$ for $m$ and $n$
        integer.  This sum guarantees that the function $K\of{x,y}$ is
        periodic in its first argument.

The coefficients $U_{i j} = \pi \, U\of{\pi \, \sqof{z_j - z_i} / L} / L$ are given by
\begin{widetext}
\bea
	\frac{\pi}{L} \, U\of{\frac{\pi}{L} \, z} = \frac{\pi}{L} \, W\of{\frac{\pi}{L} \, z} + \frac{1}{N-2} \, \sqof{\left. \frac{d}{dx} P\of{x, -z} \right|_0 + \lim_{R \rightarrow 0} \sum_{0 < \abs{z_0} \leq R} \, \frac{P\of{z_0, -z}}{z_0}},
	\label{eq:Udef}
\eea
where $W\of{z}$ is the periodic extension of $1/z$ to the
torus~\cite{laughlin89ap163} and also related to the logarithmic
derivatives of the theta functions: \bea \frac{\pi}{L} \,
W\of{\frac{\pi}{L} \, z} = \frac{d}{dz} \, \ln
\vartheta\of{\frac{\pi}{L} \, z} + \frac{\pi}{L} \, \frac{z - z^*}{2
  \, L} \, .
	\label{eq:Wdef}
\eea

The function $P\of{x,y}$ is given by
\bea
	P\of{x,y} = \lim_{R \rightarrow \infty} \sum_{0 \leq \abs{z_0 - y} \leq R} \frac{\mathrm{Co}\of{\frac{\pi}{2 \, L} \, \sqof{z_0 - y}}}{\mathrm{Co}\of{\frac{\pi}{2 \, L} \, \sqof{x - \of{y - z_0}}}} \, \frac{e^{-\frac{\pi}{L^2} \, \abs{z_0 - y}^2}}{n\of{y}},
\eea
\end{widetext}
where $\mathrm{Co}\of{x} = \cos x + \cosh x$ and where $n\of{y}$ is a
normalization factor chosen such that $P\of{0,y} = 1$ which entails
the choice \bea n\of{z} = \vartheta_3\of{\left. \frac{\pi}{L} \,
    \mathrm{Re}\sqof{z} \right| i } \, \vartheta_3\of{\left.
    \frac{\pi}{L} \, \mathrm{Im}{\sqof{z}} \right| i } \, .  \eea
While the form of the coefficients as given by
Eqs.~\ref{eq:Kdef}--\ref{eq:Wdef} are essential for forming a
Hamiltonian that stabilizes the CSL, there is significant freedom in
how one chooses the function $P\of{x,y}$.  The only requirements are
that it be a periodic function of $y$, fall off faster than $1/x$ with
increasing $x$, and be analytic apart from first-order
poles that occur at the coincidence of the two arguments: $x = y$.  It
is straightforward to show that $U\of{z}$ is an odd function; this in
turn guarantees that $\sum_i U_{ij} = 0$ and lets this sum be dropped,
as was done in writing down Eq.~\ref{eq:omegaexplicit}.

\section{Proof of Solution}
\label{sec:proof}

In order to prove that either of the Hamiltonians given in
Eqs.~\ref{eq:hvec} and \ref{eq:htense} are true parent
Hamiltonians for the chiral spin liquid, we must demonstrate that
$\omega_j \, \ket{\psi} = \omega_i \, \ket{\psi}$ which we will
demonstrate by first showing that \bea \bra{z_1 \cdots z_M} \omega_j
\, \ket{\psi} = f\of{\Zcom} \, \braket{z_1 \cdots z_M}{\psi} \, ,
	\label{eq:omegaaction}
        \eea where $f\of{\Zcom}$ is a function only of the center of
        mass: $\Zcom = \sum_{i=1}^M z_i$.  This identity in turn
        follows from the fact that \bea
	\frac{\bra{z_1 \cdots z_M} \omega_j^+ \ket{\psi}}{\braket{z_1 \cdots z_M}{\psi}} = \left\{ \begin{array}{ll} f\of{\mathcal{Z}} & z_j \in \cuof{z_1 \cdots z_M} \\
            0 & \mathrm{otherwise} \, ,
	\end{array} \right. 
	\label{eq:omegaplusaction}
        \eea and the result that the function $f\of{\Zcom}$ is both
        odd and periodic.  To see this, recall that one can write
        $\omega_j^- = U \, \omega_j^+ \, U^{\dagger}$ where $U$
        performs the $\pi/2$ rotation about the $x$-axis as discussed
        in Section~\ref{sec:destruction} above.  The CSL ground state
        is invariant under such a rotation so that \bea
	\bra{z_1 \cdots z_M} \omega_j^+ \ket{\psi} & = & \bra{z_1 \cdots z_M} U^{\dagger} \omega_j^- U \ket{\psi} \nonumber \\
	& = & \bra{w_1 \cdots w_M} \omega_j^- \ket{\psi}, \eea where
        $\cuof{w_i}$, the locations of the down spins on the lattice,
        is the complement of $\cuof{z_i}$.  It then follows from
        Eq.~\ref{eq:omegaplusaction} that \bea
	\frac{\bra{z_1 \cdots z_M} \omega_j^- \ket{\psi}}{\braket{z_1 \cdots z_M}{\psi}} = \left\{ \begin{array}{ll} 0 & z_j \in \cuof{z_1 \cdots z_M}  \\
            f\of{\mathcal{W}} & \mathrm{otherwise} \, . \end{array}
        \right.
	\label{eq:omegaminusaction}
        \eea Assuming that the origin of the lattice is chosen such
        that the sites occupy positions $z_i = \of{\ell + i \, m}$ for
        $\ell$ and $m$ integer, it is straightforward to show that
        \bea \mathcal{Z} + \mathcal{W} = \frac{L \, \of{L - 1}}{2} \,
        \of{1 + i} \, L, \eea and since $L$ is even it follows that the
        sum of $\mathcal{Z}$ and $\mathcal{W}$ is equivalent to a
        translation of the lattice $z_0$.  Because the function
        $f\of{\mathcal{Z}}$ is periodic and odd, both properties will
        be shown below, it immediately follows that $f\of{\mathcal{W}}
        = f\of{z_0 - \mathcal{Z}} = - f\of{\mathcal{Z}}$.  Combining
        this fact with Eq.~\ref{eq:omegaminusaction} completes
        the proof that Eq.~\ref{eq:omegaplusaction} entails
        Eq.~\ref{eq:omegaaction}.

\subsection{Action of $T_j$}

In order to prove Eq.~\ref{eq:omegaplusaction}, we first consider
the off-diagonal terms in the operator $\omega_j^+$ which come from
$T_j$ defined in Eq.~\ref{eq:tjdef}.  We consider a general
element of the vector $T_j \, \ket{\psi}$:
\begin{widetext}
  \bea \bra{z_1 \cdots z_M} T_j \ket{\psi} = \frac{1}{2} \sum_{i,k
    \neq j}' K_{i j k} \left\langle{z_1 \cdots z_M}\left| S_j^+ \,
      S_k^- \, \of{\frac{1}{2} + S_i^z} \right|{\psi}\right\rangle \,
  .  \eea The element is clearly zero unless $z_j \in \cuof{z_1 \cdots
    z_M}$.  When this is satisfied, acting onto the bra on the
  right-hand side of the equation with the spin operators wipes out
  the matrix element unless $z_i \in \cuof{z_1 \cdots z_M}$ and
  replaces $z_j$ with $z_k$: \bea \bra{z_1 \cdots z_M} T_j \ket{\psi}
  = \frac{1}{2} \sum_{i \neq j}^M \sum_{k \neq j}^{N} K_{i j
    k} \braket{z_1\cdots z_{j-1} \, z_k \, z_{j+1} \cdots z_M}{\psi}
  \, .  \eea The upper limit of $M=N/2$ (rather than $N$) on the
  sum on $i$ indicates that $z_i$ must be a member of the up-spins.
  Rewriting $K_{i j k} = K\of{z_k - z_j, z_i - z_j}$ and defining $z =
  z_k - z_j$, this may be rewritten as \bea \bra{z_1 \cdots z_M} T_j
  \ket{\psi} = \frac{1}{2} \sum_{i \neq j}^M \sum_{z \neq 0}
  K\of{z,z_i-z_j} \, \braket{z_1 \cdots z_j + z \cdots z_M}{\psi} \, .
  \eea Using the definition of the coefficient $K$ from
  Eq.~\ref{eq:Kdef}, this can be rewritten as \bea \bra{z_1
    \cdots z_M} T_j \ket{\psi} = \frac{1}{N-2} \sum_{i \neq j}^M
  \sum_{z \neq 0} \of{\lim_{R \rightarrow \infty} \sum_{0 \leq z_0 <
      R} \frac{P\of{z - z_0, z_i - z_j}}{z - z_0}} \braket{z_1 \cdots
    z_j + z \cdots z_M}{\psi} \, .  \eea Since the wave function
  itself is periodic, the two sums over $z$ and over $z_0$ may be
  combined into a single sum that runs over the entire infinite
  lattice for which we use the variable $x = z - z_0$.  However, since
  the point $z = 0$ is missing from the original sum, all of its
  images in the infinite lattice will be missing from the second sum
  and this must be subtracted off, giving \bea
  \bra{z_1 \cdots z_M} T_j \ket{\psi} & = & \frac{1}{N-2} \sum_{i \neq j}^M \of{ \lim_{R \rightarrow \infty} \sum_{0 < \abs{x} < R} \frac{P\of{x,z_i - z_j}}{x} \, \braket{z_1 \cdots z_j + x \cdots z_M}{\psi}} \nonumber \\
  & & - \frac{1}{N-2} \sum_{i \neq j}^M \sum_{z_0} \frac{P\of{-z_0,
      z_i - z_j}}{-z_0} \, \braket{z_1 \cdots z_M}{\psi} \, .
	\label{eq:withratio}
        \eea Dividing both sides of the equation by $\braket{z_1
          \cdots z_M}{\psi}$ and rewriting the ratio of elements in
        terms of the analytic function of $x$, $A\of{x}$ given in
        Appendix~\ref{app:analytic} yields \bea \frac{\bra{z_1 \cdots
            z_M} T_j \ket{\psi}}{\braket{z_1 \cdots z_M}{\psi}} = -
        \frac{1}{N-2} \sum_{i \neq j}^M \lim_{R \rightarrow \infty}
        \sum_{0 < \abs{x} < R} \frac{P\of{x, z_i - z_j}}{x} \, A\of{x}
        \, G\of{x} \, e^{-\frac{\pi}{2} \, \abs{x}^2} - \frac{1}{N-2}
        \sum_{i \neq j}^M \sum_{z_0} \frac{P\of{z_0, z_i - z_j}}{z_0}
        \, .
	\label{eq:readytosum}
\eea
Note that $A\of{x}$ is an analytic function only of $x$, and not of the remaining $\cuof{z_i}$ on which it also depends.

The first sum in Eq.~\ref{eq:readytosum} may be evaluated with the
corollary to the Singlet Sum Rule, Eq.~\ref{eq:corollary}. A derivation
of the sum rule and the necessary corollary is given in
Appendix~\ref{app:sumrule}.  The function $P\of{x,y}$ falls off
exponentially with increasing $x$ while the quantity $A\of{x} \,
G\of{x} \, e^{-\frac{\pi}{2} \, \abs{x}^2}$ is essentially constant
due to the periodicity of the wave function.  This guarantees that the
sum is absolutely convergent and the sum rule may be applied.
Additionally, the product $P\of{x, z_i - z_j} \, A\of{x}$ is itself an
analytic function of $x$.  As a function of $x$, the function
$P\of{x,z_i - z_j}$ necessarily has poles. However, these occur
when $x = z_i - z_j$ and the function $A\of{x}$ has second-order
zeroes at these locations since this corresponds to a coincidence of
up-spins.  Since the product is analytic and the sum is absolutely
convergent, the singlet sum rule may be applied to give \bea
\frac{\bra{z_1 \cdots z_M} T_j \ket{\psi}}{\braket{z_1 \cdots
    z_M}{\psi}} = - \frac{1}{N-2} \sum_{i \neq j}^M \, \left.
  \frac{d}{dx} \, \sqof{ P\of{x, z_i - z_j} \, A\of{x} } \right|_0 -
\frac{1}{N-2} \sum_{i \neq j}^M \sum_{z_0} \, \frac{P\of{z_0, z_i -
    z_j}}{z_0} \, .  \eea Using the fact that $A\of{0} = P\of{0, z_i -
  z_j} = 1$ and the relation for $dA/dx$ given in Eq.~\ref{eq:dAdx},
this becomes \bea
\frac{\bra{z_1 \cdots z_M} T_j \ket{\psi}}{\braket{z_1 \cdots z_M}{\psi}} & = & - \frac{1}{N-2} \, \sum_{i \neq k, j}^M \left\{ \sum_{\nu = 1}^2 \, \frac{\pi}{L} W\of{\frac{\pi}{L} \, \sqof{\Zcom - Z_{\nu}}} + 2 \,  \sum_{\ell \neq j}^{M} \frac{\pi}{L} \, W\of{\frac{\pi}{L} \, \sqof{z_j - z_{\ell}} } \right. \\
& & + \left. \left. \frac{d}{dx} P\of{x, z_i - z_j} \right|_0 \right\}
- \frac{1}{N-2} \sum_{i \neq j, k}^M \sum_{z_0} \, \frac{P\of{z_0, z_i
    - z_j}}{z_0} \, .  \eea The sum on $i$ may be completed for the
terms containing the $W$ functions (picking up a factor of $M = N/2 -
1$) and this gives, renaming $\ell$ as $i$, \bea \frac{\bra{z_1 \cdots
    z_M} T_j \ket{\psi}}{\braket{z_1 \cdots z_M}{\psi}} =
f\of{\mathcal{Z}} - \sum_{i \neq j}^M \frac{\pi}{L} \,
W\of{\frac{\pi}{L} \, \sqof{z_j - z_i}} - \frac{1}{N-2} \sum_{i \neq
  j}^M \sqof{\sum_{z_0} \frac{P\of{z_0, z_i - z_j}}{z_0} + \left.
    \frac{d}{dx} P\of{x,z_i - z_j} \right|_0} \, , \eea
\end{widetext}
where \bea f\of{\mathcal{Z}} = -\frac{1}{2} \, \sum_{\nu = 1}^2
\frac{\pi}{L} \, W\of{\frac{\pi}{L} \, \sqof{\mathcal{Z} - Z_{\nu}}}
\, .  \eea The fact that $f\of{\mathcal{Z}}$ is both odd and periodic,
required for the proof of Eq.~\ref{eq:omegaaction} above, follows
from these same properties of the $W$ function.  Comparison with
Eq.~\ref{eq:Udef} shows that \bea \frac{\bra{z_1 \cdots z_M} T_j
  \ket{\psi}}{\braket{z_1 \cdots z_M}{\psi}} = f\of{\Zcom} - \sum_{i
  \neq j}^M U_{i j}
	\label{eq:finalTaction}
\eea
if $z_j$ is an element of the up-spins and zero otherwise.

\subsection{Action of $V_j$}

The action of the operator $V_j$ on the CSL ground state is
straightforward to compute.  Proceeding in an analogous manner, we have
\bea
	&&\bra{z_1 \cdots z_M} V_j \ket{\psi} = \nonumber \\
&& \sum_{i \neq j}^{N} U_{i j} \, \bra{z_1 \cdots z_M} \, \of{\frac{1}{2} + S_i^z} \, \of{\frac{1}{2} + S_j^z} \ket{\psi} \, .
\eea
The matrix element vanishes unless both $z_i$ and $z_j$ are elements
of $\cuof{z_1 \cdots z_M}$.  Therefore, the diagonal contribution to
the operator $\omega_j$ gives \bea \frac{\bra{z_1 \cdots z_M} V_j
  \ket{\psi} }{\braket{z_1 \cdots z_M}{\psi}} = \sum_{i \neq j}^M U_{i
  j}
	\label{eq:finalVaction}
        \eea if $z_j \in \cuof{z_1 \cdots z_M}$ and $0$ otherwise.
        Combining Eqs.~\ref{eq:finalVaction} and
        \ref{eq:finalTaction} yields Eq.~\ref{eq:omegaplusaction}
        and therefore proves that the chiral spin liquid is an exact
        ground state of either of the Hamiltonians in
        Eqs.~\ref{eq:hamvec} or \ref{eq:hamtense}.

\section{Kernel Sweeping Method}
\label{sec:kernelsweeping}
  To implement the Hamiltonians given in Eq.~\ref{eq:hamvec} and
  Eq.~\ref{eq:hamtense}, one has to take into account that 6-body
  terms appear in the Hamiltonians. For microscopic models containing
  many-body interactions, one must be very efficient if one hopes
  to write down the Hamiltonian in a reasonable amount of time.  For
  our Hamiltonians, this is because there are, even for a lattice with
  only $N = 16$ sites, literally thousands of terms in the Hamiltonian
  corresponding to all the different ways to choose six sites out of
  sixteen.  In contrast, a model with only two-site interactions on
  the same lattice would only have $15$ terms to compute after taking
  into account translational symmetry, even if the model had infinite
  range.  In this section, we describe an algorithm for calculating
  the Hamiltonian very efficiently, called the kernel sweeping method.

  As an example to illustrate the kernel sweeping method, we will
  consider the computation of a Heisenberg-type Hamiltonian such as
  \bea H = \sum_{i j} J_{i j} \, \Svec_i \cdot \Svec_j \, .  
  \eea 
  We work in an $S^z$ basis and label the states by a binary number
  where up-spins are treated as $1$'s and down-spins are treated as
  $0$'s.  We first note that since this is a two-site interaction, in
  order to implement this model all we really need to know is how the
  operator $\Svec_i \cdot \Svec_j$ acts on the four-dimensional basis
  $\ket{s_i \, s_j}$.  This action may be summarized as 

\bea
		\begin{array}{cccc|c}
	\bra{\down \down} & \bra{\down \up} & \bra{\up \down} & \bra{\up \up} \\ \hline
	1/4 & 0 & 0 & 0 & \ket{\down \down} \\
	0 & -1/4 & 1/2 & 0 & \ket{\down \up} \\
	0 & 1/2 & -1/4 & 0 & \ket{\up \down} \\
	0 & 0 & 0 & 1/4 & \ket{\up \up}
	\end{array} \, 
\eea
where the table format shows the order of the basis vectors.  It is
only necessary to compute this matrix once at the beginning of running
the code.  One stores this matrix as a set of rules
\bea
	R = \cuof{ \sqof{\cuof{s}_m,\cuof{s}_n} \rightarrow \Omega_{m n} }
\eea
where the $\cuof{s}_m$ and the $\cuof{s}_n$ are a binary shorthand for
the states in this two-dimensional basis and $\Omega_{mn}$ are the
elements in the matrix. In this example we would have 
\bea R &=& \cuof{
  \sqof{00,00} \rightarrow \frac{1}{4}, \sqof{01,01} \rightarrow
  -\frac{1}{4}, \sqof{01,10} \rightarrow \frac{1}{2}, \right. \nonumber\\
&& \left. \sqof{10,10}
  \rightarrow - \frac{1}{4}, \sqof{11,11} \rightarrow \frac{1}{2}}
\eea 
where, since we are dealing with a Hermitian operator, we only
need to include the upper triangle.  The extension of this array to a
$p$-site operator is straightforward; in that case one must consider
the action of the operator on a $2^p$-dimensional basis.  Therefore,
the corresponding operator for the chiral spin liquid Hamiltonian
given in Equation~\ref{eq:hamvec} is 64-dimensional.

The code next loops over all possible values of $i$ and $j$ and does
the following.  First it computes
\bea
	R_{ij} &=& \cuof{ \sqof{ \cuof{s}_m \cdot \cuof{2^{i-1},2^{j-1}}, \cuof{s}_n \cdot \cuof{2^{i-1},2^{j-1}}} \right. \nonumber \\
 && \left. \rightarrow  J_{i j} \, \Omega_{m n}} \, .
\eea
All this means is to compute the contribution of the two spins at
sites $i$ and $j$ to the binary number that will label the entire
state.  For our example, assuming that we are at a point in the loop
where $i = 3$ and $j = 7$, this gives
\bea
	R_{3 7} &=& \cuof{ \sqof{0,0} \rightarrow \frac{J_{37}}{4}, \sqof{2^{6},2^{6}} \rightarrow -\frac{J_{37}}{4}, \sqof{2^{6},2^{2}} \rightarrow \frac{J_{37}}{2}, \right. \nonumber \\
&& \hspace{-4pt} \left. \sqof{2^{2},2^{2}} \rightarrow -\frac{J_{37}}{4}, \sqof{2^2 + 2^6, 2^2 + 2^6} \rightarrow \frac{J_{37}}{4}} . 
\eea
The code next computes the contributions to the binary numbers
labeling the states from all the sites that are not involved in the
interaction.  There are $2^{N-p}$ of these and for our two-site
example this list is
\bea
	B_{i j} = \cuof{\sum_{l \neq i,j}^N s_l \, 2^{l - 1}} \, .
	\label{eq:Bs}
\eea
Finally, one updates the Hamiltonian according to 
\bea
	H = H + R_{i j} \otimes B_{i j}  
\eea
where the addition means to add the matrix defined by these rules and
the generalized outer product means
\begin{widetext}
\bea
	R_{i j} \otimes B_{i j}  = \cuof{\sqof{\cuof{s}_m \cdot \cuof{2^{i-1},2^{j-1}} + b, \cuof{s}_n \cdot \cuof{2^{i-1},2^{j-1}} + b} \rightarrow  J_{i j} \, \Omega_{m n}}
\eea
\end{widetext}
for $b$ an element of $B_{i j}$.  In this way one may construct the
Hamiltonian extremely quickly since all the steps involve list
operations and there is only a single loop over the $N$ choose $2$
ways to pick the sites $i$ and $j$. (In practice, one uses
translational invariance to fix $i = 1$ and, for a two-site operator
as in this example, the loop is then over the $N-1$ ways to choose the
remaining site.)

Let us now work this out explicitly for the 
vector Hamilton
operator Eq.~\ref{eq:hamvec}. Setting $\mathbf{D}_{ij}=\Omegavec_i -
\Omegavec_j$, we split up the Hamiltonian into
\begin{equation}
\label{hamprac}
  H =\sum_{\expect{i j}} \Omegavec_{ij}^{z,\dagger}\Omegavec_{ij}^{z}+\frac{1}{2}\left(\Omegavec_{ij}^{+,\dagger}\Omegavec_{ij}^{+}+\Omegavec_{ij}^{-,\dagger}\Omegavec_{ij}^{-}\right),
\end{equation}
where the $z$-component as well as the ladder components of the vector
operators can be written out in terms of spin operators $S^z$,
$S^+=S^x+i S^y$, and $S^-=S^x-i S^y$. As the treatment is very
similar, we constrain our attention to the contribution
$\sum_{\expect{i j}}\Omegavec_{ij}^{+,\dagger}\Omegavec_{ij}^{+}$,
where for clarity we again write out the $+$ ladder operator
explicitly:
\begin{widetext}
  \begin{eqnarray} 
    \Omegavec_j^+ &=& \sum_{i, k \neq j}' K_{ijk} \, \sqof{ \frac{1}{4
        \, i} \, (S_j^zS_k^+ - S_j^+S_k^z) + \frac{4}{5} \,
      \of{\Svec_j \cdot \Svec_k} \, S_i^+ - \frac{1}{5} \, \of{\Svec_k
        \cdot \Svec_i} \, S_j^+ - \frac{1}{5} \, \of{\Svec_i \cdot
        \Svec_j} \, S_k^+}
+ \sum_{i \neq j} U_{ij} \, S_i^+  \, .
	\label{eq:omegapcomp}
\end{eqnarray}
\end{widetext}

Using the notation analogous to Eq.~\ref{eq:Bs}
\bea
	B^{\text{k}}_{i j  k} = \cuof{\sum_{l \neq i,j,k}^N s_l \, 2^{l - 1}} \, \;\;\;\; B^{\text{u}}_{i} = \cuof{\sum_{l \neq i}^N s_l \, 2^{l - 1}},
\eea
we can write
\begin{equation}
\Omegavec_j^+=\sum_{\substack{i\\ i\ne j}}\left[\sum_{\substack{k\\ k\ne i,j}} R^{\text{k}}_{i j k } \otimes B^{\text{k}}_{i j k} + R^{\text{u}}_{i} \otimes B^{\text{u}}_{i}\right],
\end{equation}
where $R^{\text{k}}_{i j k }$ and $R^{\text{u}}_{i}$ relate to the
first and second sum of Eq.~\ref{eq:omegapcomp}, respectively. Given
these 3-body operators in above notation, the total 6-body interaction
can be conveniently computed. The implementation of the tensor
Hamilton operator Eq.~\ref{eq:hamtense} is completely analogous.

\section{Numerical Confirmation}
\label{sec:numerics}
\begin{figure}[b]
\centerline{\includegraphics[scale=0.33]{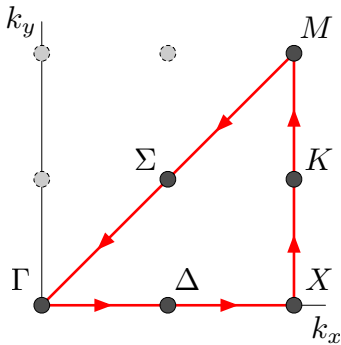}}
\caption{A plot of the symmetry points in the first Brillouin zone.
  The arrows show the path taken in plotting the energy spectra in
  Fig.~\ref{fighspec}, starting
  from the origin at $\Gamma = \of{0,0}$.}
\label{fig:lattice}
\end{figure}

\begin{figure}[t]
\includegraphics[width=0.9\linewidth]{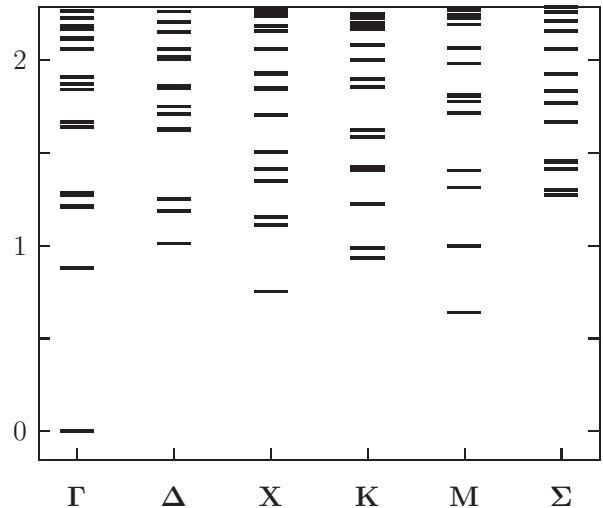}
\caption{Low energy spectrum of the
  Hamiltonian Eq.~\ref{eq:hamvec}, scaled down to order of unity.
  There are two $E=0$ eigenvalues at the $\Gamma$ point.}
\label{fighspec}
\end{figure}

Using the method outlined in Section~\ref{sec:kernelsweeping} above,
the models in Eq.~\ref{eq:hamvec} and Eq.~\ref{eq:hamtense} have been
solved by exact diagonalization on 16-site lattices with periodic
boundary conditions.
We start by considering the vector Hamiltonian given by
Eq.~\ref{eq:hamvec}. The spectrum is shown in Fig.~\ref{fighspec}; the
points in the Brillouin zone which label the axis of this figure are
shown in Fig.~\ref{fig:lattice}.
We find the spectrum to be positive semi-definite, with a
doubly-degenerate zero-energy state at the $\Gamma$ point. The rest of
the spectrum is well separated from the ground state by a gap that is
substantial and we believe not due to finite size effects in the
calculation.  This claim is based on the fact that it exceeds the
finite size level splitting of the spectrum by a factor of $\sim 15$.
The presence of a gap is expected between the chiral spin liquid
ground state and what should be a two-spinon excited state. The spinon
excitations of this model will be addressed in future work.

We now discuss the two orthogonal zero-energy eigenstates.  For
comparison, we construct the CSL state Eq.~\ref{eq:wavefunction}
explicitly and find a two-dimensional subspace of functions with the
center of mass variable being treated as an external parameter. We
have computed the overlap of the Hamiltonian ground state subspace and
the CSL subspace and find that they match perfectly. Therefore, the
ground state of this Hamiltonian is indeed the two-fold degenerate CSL
state. Additionally, we have {\it only two} zero-energy states, by
which follows that the CSL state is the {\it only} ground state of the
model, a statement which cannot be achieved analytically.

For the tensor Hamiltonian, however, we find that the zero-energy
subspace is massively degenerate. It of course contains the CSL, in
accord with the analytical proof, but also many additional states.
While the restriction to small system sizes prevents us from studying
the thermodynamic limit precisely, our numerical findings indicate
that the Hamiltonian Eq.~\ref{eq:hamtense} does not stabilize the CSL
state as the unique ground state, which thus singles out the model in
Eq.~\ref{eq:hamvec} to be subject of further study.
\section{Conclusion}
\label{sec:conclusion}
In this work we have shown a method for constructing parent
Hamiltonians for the chiral spin liquid. We have computed the spectra
of the Hamiltonians by use of a Kernel-sweeping method in exact
diagonalization. There, for the Hamiltonian operator composed of the
spherical vector component of the CSL destruction operator, we observe
that the CSL states are the only ground states of the model. We
conclude that this model is a promising candidate to also study the
elementary excitations of the model, \ie spinons, and many other
questions in the field of two-dimensional fractionalization of quantum
numbers in spin systems.

\begin{acknowledgments}
  RT was supported by a PhD scholarship from the Studienstiftung des
  deutschen Volkes; DS acknowledges support from the Research
  Corporation under grant CC6682.  We would like to thank J.S.
  Franklin, R. Crandall, and R.B.  Laughlin for many useful
  discussions.
\end{acknowledgments}

\begin{appendix}
\section{Tensor decomposition}
\label{app:decomp}

The operators $\omega_j$ introduced in Section~\ref{sec:destruction}
may be decomposed into irreducible spherical tensors of ranks $1$ and
$3$.  We write these irreducible operators as $T^{q}_m$; $q$ and $m$
correspond to angular momentum and its $z$ component respectively. We
wish to write $\omega = \sum c_q \, T^q$, where $T^q$ is the collection
of all operators which transform as a spherical tensor of rank $q$.
Here we have suppressed the site index on the operator $\omega$.

The operator in Eq.~\ref{eq:omegaexplicit} that is not manifestly
the component of a vector is $S_i^z \, S_j^z \, S_k^z$, which is a
component of a third-rank Cartesian tensor.  In order to keep the
notation manageable, we start by considering the direct product of two
operators $U$ and $V$ with angular momentum $j_1$ and $j_2$
respectively.  An element in the direct product space of these
operators may be written as \bea U^{j_1}_{m_1} \, V^{j_2}_{m_2} =
\sum_{j_{12} = \abs{j_2 - j_1}}^{j_1 + j_2} \, \sum_{m_{12} =
  -j_{12}}^{j_{12}} \, \cg{j_1}{m_1}{j_2}{m_2}{j_{12}}{m_{12}} \,
T^{j_{12}}_{m_{12}}
	\label{eq:addition1}
        \eea in terms of irreducible spherical tensors
        $T_{m_{12}}^{j_{12}}$ carrying angular momentum $j_{12}$ with
        $z$-component $m_{12} = m_1 + m_2$.
        Eq.~\ref{eq:addition1} may be inverted to give \bea
        T_{m_{12}}^{j_{12}} = \sum_{m_1 = -j_1}^{j_1} \sum_{m_2 = -
          j_2}^{j_2} \cg{j_1}{m_1}{j_2}{m_2}{j_{12}}{m_{12}} \,
        U^{j_1}_{m_1} \, V^{j_2}_{m_2} \, .
	\label{eq:addition2}
\eea

Using these equations, one may construct corresponding expressions for
the product of three vector operators by applying
Eq.~\ref{eq:addition1} twice:
\begin{widetext}
  \bea
  U_{m_1}^{j_1} \, V_{m_2}^{j_2} \, W_{m_3}^{j_3} & = & \sum_{j_{12} = - \abs{j_1 - j_2}}^{j_1 + j_2} \sum_{m_{12} = - j_{12}}^{j_{12}} \cg{j_1}{m_1}{j_2}{m_2}{j_{12}}{m_{12}} \, T_{m_{12}}^{j_{12}} \, W^{j_3}_{m_3} \nonumber \\
  & = & \sum_{j_{12} = - \abs{j_1 - j_2}}^{j_1 + j_2} \sum_{m_{12} = -
    j_{12}}^{j_{12}} \, \cg{j_1}{m_1}{j_2}{m_2}{j_{12}}{m_{12}} \,
  \sum_{j = \abs{j_{12} - j_3}}^{j_{12} + j_3} \sum_{m = - j}^{j}
  \cg{j_{12}}{m_{12}}{j_3}{m_3}{j}{m} \, T^{j \, \of{j_{12}}}_m \, .
  \eea The second superscript on the tensor $T$ in the last line
  distinguishes between the different tensors of the same rank that
  appear when combining three vector operators; since $1 \otimes 1
  \otimes 1 = 3 \oplus 2 \oplus 2 \oplus 1 \oplus 1 \oplus 1 \oplus
  0$, there are two rank-2 spherical tensors and three vector
  operators that can be formed.  For the case of interest $m_1 = m_2 =
  m_3 = 0$ and $j_1 = j_2 = j_3 = 1$, the expression reduces to
  \bea U^z \, V^z \, W^z = \sum_{j_{12} = 0}^2 \sum_{j = \abs{j_{12} -
      1}}^{j_{12} + 1} \cg{1}{0}{1}{0}{j_{12}}{0} \,
  \cg{j_{12}}{0}{1}{0}{j}{0} \, T_0^{j \, \of{j_{12}}} = -
  \frac{1}{\sqrt{3}} \, T^{1 \, \of{0}}_0 - \frac{2}{\sqrt{15}} \,
  T^{1 \, \of{2}}_0 + \sqrt{\frac{2}{5}} \, T^{3}_0 \, ,
	\label{eq:breakdown}
        \eea which shows that the operator contains only vector and
        rank-3 tensor components, but no scalar or rank-2 tensor
        components.  Note that the second index on the rank-3 tensor
        has been suppressed since the construction of this object is
        unambiguous.

Applying Eq.~\ref{eq:addition2} twice, the rank-$3$ tensor component is
\bea
	T^{3
	}_0 & = & \sum_{m_{12} = - 2}^2 \sum_{m_3 = -1}^1 \cg{2}{m_{1 2}}{1}{m_3}{3}{0} \, T^{2}_{m_{12}} \, W^1_{m_3}  \\
	& = & \sum_{m_1,m_2,m_3 = -1}^1 \, \cg{2}{-m_3}{1}{m_3}{3}{0} \, \cg{1}{m_1}{1}{m_2}{2}{-m_3} U^1_{m_1} \, V^1_{m_2} \, W^1_{m_3} \\
	& = & \frac{5 \, U^z \, V^z \, W^z - \of{\Uvec \cdot \Vvec} \, W^z - \of{\Vvec \cdot \Wvec} \, U^z - \of{\Wvec \cdot \Uvec} \, V^z}{\sqrt{10}} \, ,
	\label{eq:threetensor}
        \eea where we have used the fact that the dot product is
        $\mathbf{U} \cdot \mathbf{V} = \sum_m \of{-1}^m U^1_m \,
        V^1_{-m}$ in the spherical representation. A similar
        construction can be used to find the vector operator or, one
        may note from Eqs.~\ref{eq:breakdown} and
        \ref{eq:threetensor} that the vector component is equivalent
        to \bea U^z \, V^z \, W^z - \sqrt{\frac{2}{5}} \,
        T_0^{3
	} = \frac{\of{\Uvec \cdot \Vvec} \, W^z + \of{\Vvec \cdot
            \Wvec} \, U^z + \of{\Wvec \cdot \Uvec} \, V^z}{5}
	\label{eq:vectorpart}
\eea
as used in writing down Eq.~\ref{eq:omegavecdef}.

Construction of the remaining ($x$ and $y$) components of the vector
operator in Eq.~\ref{eq:vectorpart} is straightforward since one
merely replaces $z$ with either $x$ or $y$.  In order to construct the
remaining six components of the rank-3 tensor operator one simply
applies Eq.~\ref{eq:addition2} twice without specifying $m = 0$:
\bea T_m^3 = \sum_{m_1, m_2, m_3 = -1}^1\cg{2}{m-m_3}{1}{m_3}{3}{m} \,
\cg{1}{m_1}{1}{m_2}{2}{m-m_3} \, U^1_{m_1} \, V^1_{m_2} \, W^1_{m_3}
\, .  \eea The explicit form of these components are \bea
T_1^3 & = & - \frac{1}{2 \, \sqrt{30}} \, \sqof{ \of{5 \, V^z \, W^z - \Vvec \cdot \Wvec} \, U^+ + \of{5 \, U^z \, W^z - \Uvec \cdot \Wvec} \, V^+ + \of{5 \, U^z \, V^z - \Uvec \cdot \Vvec} \, W^+  }  \\
T_2^3 & = & \frac{1}{2 \, \sqrt{3}} \, \sqof{U^+ \, V^+ \, W^z + U^+ \, V^z \, W^+ + U^z \, V^+ \, W^+} \\
T_3^3 & = & - \frac{1}{2 \, \sqrt{2}} \, U^+ \, V^+ \, W^+ \, , \eea
with the remaining three components obtained from $T^q_{-m} =
\of{-1}^m \, \of{T^q_m}^{\dagger}$.
\vspace{3pt}
\end{widetext}

\section{Sum Rule}
\label{app:sumrule}

The sum rule used in Section~\ref{sec:proof}, on which the proof that
$\omega$ destroys the ground state hinges, is given by \bea \lim_{R
  \rightarrow \infty} \sum_{0 \leq \abs{z} < R} G\of{z} \, z^n \,
e^{-\frac{\pi}{2} \, \abs{z}^2} = 0 \, .
	\label{eq:sumrule}
        \eea The sum rule has been first stated in a mathematical
        framework of coherent state systems by Perelomov and has later
        been re-derived by
        Laughlin~\cite{perelomov71tmp156,laughlin89ap163}; in this
        appendix, we show how to obtain the sum rule by application of
        Jacobi's imaginary transformation.  We first consider the
        related sum \bea F\of{c} = \lim_{R \rightarrow 0} \sum_{0 \leq
          \abs{z} < R} G\of{z} \, \exp\sqof{\frac{1}{2} \, c \, z -
          \frac{\pi}{2} \, \abs{z}^2} \, .
	\label{eq:bigFdef}
        \eea In order to prove the sum rule in
        Eq.~\ref{eq:sumrule}, we will first show that $F\of{c} =
        0$ for any value of the parameter $c$ and then use this to
        prove Eq.~\ref{eq:sumrule} by taking derivatives of the
        function $F\of{c}$.

        In order to show that $F\of{c} = 0$, we use the gauge function
        $G\of{z}$ to write Eq.~\ref{eq:bigFdef} as two sums, one
        over the entire lattice and one over the points $z'$ on the
        lattice for which $G\of{z'} = - 1$.  As shown in
        Figure~\ref{fig:lattice}, these sites defines a sublattice
        with twice the original lattice spacing.  \bea F\of{c} =
        \sum_{z} e^{\frac{1}{2} \, c \, z - \frac{\pi}{2} \,
          \abs{z}^2} - 2 \, \sum_{z'} e^{\frac{1}{2} \, c \, z' -
          \frac{\pi}{2} \, \abs{z'}^2} \, .  \eea Setting $z' = 2 \,
        z$ we can write this as \bea F\of{c} = \sum_{z} e^{\frac{1}{2}
          \, c \, z - \frac{\pi}{2} \, \abs{z}^2} - 2 \, \sum_z e^{c
          \, z - 2 \, \pi \, \abs{z}^2} \, , \eea where both sums now
        run over the entire lattice.  Writing $z = x + i \, y$ this
        function can be factored into four sums over the integers $x$
        and $y$: \bea
	F\of{c} & = & \of{\sum_x e^{\frac{1}{2} \, \of{c \, x - \pi \, x^2}}} \, \of{\sum_y e^{\frac{1}{2} \of{i \, c \, y - \pi \, y^2}}} \nonumber\\
	& & - 2 \, \of{\sum_x e^{c \, x - 2 \, \pi \, x^2}} \,
        \of{\sum_y e^{i \, c \, y - 2 \, \pi \, y^2}} \, .  \eea In
        terms of the third Jacobi theta
        function\cite{AbramowitzStegun65} \bea \theta_3\of{z | \tau} =
        \sum_{n = - \infty}^{\infty} e^{i \, \pi \, n^2 \, \tau} \,
        e^{2 \, i \, n \, z} \, , \eea this function may be recast as
        \bea
	F\of{c} & = &  \vartheta_3\of{\left.\!- i \, \frac{c}{4}  \right| \frac{i}{2}} \, \vartheta_3\of{\left.\! \frac{c}{4} \right| \frac{i}{2}} \nonumber \\
	& & - 2 \, \vartheta_3\of{\left.\!- i \, \frac{c}{2} \right| 2
          \, i} \, \vartheta_3\of{\left.\! \frac{c}{2} \right| 2 \, i}
        \, .
	\label{eq:thetatheta}
\eea

The fact that the two terms in this expression precisely cancel is a
result of Jacobi's imaginary transformation~\cite{Whittaker-58}, \bea
\theta_3\of{z|\tau} = \frac{1}{\sqrt{-i \, \tau}} \, e^{z^2 / i \, \pi
  \, \tau} \, \vartheta_3\of{\!\left.\pm \frac{z}{\tau} \right| -
  \frac{1}{\tau}} \, , \eea and the fact that the third Jacobi theta
function is even.  Application of this identity to either product of
theta functions in Eq.~\ref{eq:thetatheta} shows that the two terms
precisely cancel, proving that $F\of{c} = 0$.  This in turn proves the
$n = 0$ case of Eq.~\ref{eq:sumrule} by simply setting $c = 0$.  The
other instances of the sum rule are obtained by noting that \bea
\frac{1}{m!} \frac{d^m}{d c^m} F\of{c} = \lim_{R \rightarrow \infty}
\sum_{0 \leq \abs{z} < R} G\of{z} \, z^m \, e^{c \, z} \, e^{-
  \frac{\pi}{2} \, \abs{z}^2} \, .  \eea Since $F\of{c}$ is $0$ for
all values of $c$, setting $c = 0$ in the above expression gives the
desired result in Eq.~\ref{eq:sumrule}. It should be noted that this
proof can be generalized for arbitrary lattices. However, for lattice
structures with non-orthogonal vectors spanning the unit cell, a
decomposition into Jacobi Theta functions is not possible anymore and
to follow the above line of proof one has to apply the generalized
Liouville theorem for two-dimensional Riemann theta
functions~\cite{Mumford83}.

\subsubsection{Corollary to Sum Rule}

We now consider the case where we wish to evaluate a sum of the form
\bea
	\lim_{R \rightarrow \infty} \sum_{0 < \abs{z} < R} \, \frac{1}{z} \, A\of{z} \, G\of{z} \, e^{-\frac{\pi}{2} \, \abs{z}^2}
	\label{eq:sumanalytic}
        \eea where $A\of{z}$ is an analytic function of $z$.  Since it
        is analytic, we can expand the function $A\of{z}$ in a Taylor
        series: \bea A\of{z} = \sum_{\ell = 0}^{\infty}
        \frac{1}{\ell!} \, \left. \frac{d^{\ell} A}{d z^{\ell}}
        \right|_0 \, z^{\ell} \eea and, so long as the sum in
        Eq.~\ref{eq:sumanalytic} is absolutely convergent, we can
        interchange the order of the two infinite sums to obtain \bea
        \sum_{\ell} \, \frac{1}{\ell!} \, \left. \frac{d^{\ell} A}{d
            z^{\ell}} \right|_0 \, \of{\lim_{R \rightarrow \infty}
          \sum_{0 < \abs{z} < R} z^{\ell - 1} \, G\of{z} \,
          e^{-\frac{\pi}{2} \, \abs{z}^2}} \, .  \eea All terms for
        which $\ell > 2$ immediately vanish from the interior sum due
        to the sum rule in Eq.~\ref{eq:sumrule}.  The term with
        $\ell = 0$ also vanishes because in that case the summand is
        an odd function summed over the entire lattice.  Finally the
        term with $\ell = 1$ can be evaluated using the sum rule and
        is simply the negative of the value of the summand at $z = 0$
        (which is not included in this sum but is included in
        Eq.~\ref{eq:sumrule}). Therefore, so long as $A\of{z}$ is
        chosen so that the sum itself is absolutely convergent, \bea
        \lim_{R \rightarrow \infty} \sum_{0 < \abs{z} < R} \,
        \frac{1}{z} \, A\of{z} \, G\of{z} \, e^{-\frac{\pi}{2} \,
          \abs{z}^2} = \left. \frac{d A}{dz} \right|_0 \, .
	\label{eq:corollary}
\eea

\section{The function $A\of{z}$}
\label{app:analytic}

The ratio of wave function coefficients appearing in
Eq.~\ref{eq:withratio}, \bea \frac{\braket{z_1 \cdots z_j + x
    \cdots z_M}{\psi}}{\braket{z_1 \cdots z_j \cdots z_M}{\psi}} \, ,
\eea can be written in terms of the Gauge function $G\of{x}$, the
Gaussian $e^{-\frac{\pi}{2} \, \abs{x}^2}$, and an analytic function
of $x$, $A\of{x}$.  To see this we note that this ratio may be written
explicitly as
\begin{widetext}
\bea
	 \prod_{\nu = 1}^2 \frac{\vartheta\of{\frac{\pi}{L} \, \sqof{\Zcom + x - Z_{\nu}}}}{\vartheta\of{\frac{\pi}{L} \, \sqof{\Zcom  - Z_{\nu}}}} \, \prod_{i \neq j}^M \, \frac{\vartheta^2\of{\frac{\pi}{L} \, \sqof{z_j + x - z_i}}}{\vartheta^2\of{\frac{\pi}{L} \, \sqof{z_j - z_i}}} \, \frac{G\of{z_j + x}}{G\of{z_j} \, G\of{x}} \, \frac{e^{\frac{\pi}{2} \, \sqof{ \, \of{z_j + x}^2 - \abs{z_j + x}^2}}}{e^{\frac{\pi}{2} \, \sqof{ z_j^2 - \abs{z_j}^2}} \, e^{\frac{\pi}{2} \, \sqof{x^2 - \abs{x}^2}}} \, G\of{x} \, e^{\frac{\pi}{2} \, \of{x^2 - \abs{x}^2}} \, .
\eea
This simplifies by noting that the exponential terms obey an addition
formula \bea \frac{e^{\frac{\pi}{2} \, \sqof{ \, \of{z_j + x}^2 -
      \abs{z_j + x}^2}}}{e^{\frac{\pi}{2} \, \sqof{ z_j^2 -
      \abs{z_j}^2}} \, e^{\frac{\pi}{2} \, \sqof{ x^2 - \abs{x}^2}}} =
e^{\frac{\pi}{2} \, \sqof{ \, \of{x - x^*} \, z_j + x \, \of{z_j -
      z_j^*}}},
	\label{eq:product1}
\eea
and the Gauge function obeys an addition formula given by 
\bea
 	\frac{G\of{z_j + x}}{G\of{z_j} \, G\of{x}} = - e^{\frac{\pi}{2} \, \of{z_j^* \, x^* - z_j \, x}} \, .
	\label{eq:product2}
        \eea Since the terms involving $x^*$ cancel on multiplying the
        two expressions in Eqs.~\ref{eq:product1} and
        \ref{eq:product2}, the ratio of coefficients is \bea
        \frac{\braket{z_1 \cdots z_j + x \cdots
            z_M}{\psi}}{\braket{z_1 \cdots z_j \cdots z_M}{\psi}} = -
        A\of{x} \, G\of{x} \, e^{- \frac{\pi}{2} \, \abs{x}^2} \, , \eea
        where $A\of{x}$ is an analytic function of $x$:
\bea
	A\of{x} =  \prod_{\nu = 1}^2 \frac{\vartheta\of{\frac{\pi}{L} \, \sqof{\Zcom + x - Z_{\nu}}}}{\vartheta\of{\frac{\pi}{L} \, \sqof{\Zcom  - Z_{\nu}}}} \, \prod_{i \neq j}^M \, \frac{\vartheta^2\of{\frac{\pi}{L} \, \sqof{z_j + x - z_i}}}{\vartheta^2\of{\frac{\pi}{L} \, \sqof{z_j - z_i}}} \, e^{\frac{\pi}{2} \, \sqof{x^2 + x \, \of{z_j - z_j^*}}} \, .
	\label{eq:AFunctionDef}
        \eea The derivative of this function is given by \bea \frac{d
          A}{d x} = \cuof{\sum_{\nu = 1}^2 \frac{d}{dx} \ln
          \vartheta\of{\frac{\pi}{L} \, \sqof{\Zcom + x - Z_{\nu}}} +
          2 \, \sum_{i \neq j}^M \frac{d}{dx} \ln
          \vartheta\of{\frac{\pi}{L} \, \sqof{z_j - z_i + x}} +
          \frac{\pi}{2} \, \of{2 \, x + z_j - z_j^*} } \, A\of{x} \, .
        \eea Evaluating this at $x = 0$ and noting that $A\of{0} = 1$
        from Eq.~\ref{eq:AFunctionDef} gives \bea \left.
          \frac{dA}{dx} \right|_0 = \sum_{\nu = 1}^2 \frac{d}{d\Zcom}
        \ln \vartheta\of{\frac{\pi}{L} \, \sqof{\Zcom - Z_{\nu}}} + 2
        \, \sum_{i \neq j}^M \frac{d}{dz_j} \ln
        \vartheta\of{\frac{\pi}{L} \, \sqof{z_j - z_i}} + N \,
        \frac{\pi}{L} \, \frac{ z_j - z_j^*}{2 \, L} \, .  \eea In
        terms of the function $W\of{z}$ introduced in
        Eq.~\ref{eq:Wdef} this may be written as \bea \left.
          \frac{dA}{dx} \right|_0 = \sum_{\nu = 1}^2 \frac{\pi}{L} \,
        W\of{\frac{\pi}{L} \, \sqof{\Zcom - Z_1}} + 2 \, \sum_{i \neq
          j}^M \frac{\pi}{L} W\of{\frac{\pi}{L} \sqof{z_j - z_i}} \, .
	\label{eq:dAdx}
        \eea The final expression follows from the fact that the
        center of mass zeroes are constrained to satisfy $\sum_{\nu}
        Z_{\nu} = 0$ as pointed out in Section~\ref{sec:groundstate}.
\end{widetext}

\end{appendix}




\begin{thebibliography}{10}

\bibitem{su-79prl1698}
W.~P. Su, J.~R. Schrieffer, and A.~J. Heeger, Phys. Rev. Lett. {\bf 42},  1698
  (1979).

\bibitem{leinaas-77ncb1}
J.~M. Leinaas and J. Myrheim, Nuovo Cimento~B {\bf 37},  1  (1977).

\bibitem{wilczek82prl1144}
F. Wilczek, Phys. Rev. Lett. {\bf 48},  1144  (1982).

\bibitem{wilczek82prl957}
F. Wilczek, Phys. Rev. Lett. {\bf 49},  957  (1982).

\bibitem{laughlin83prl1395}
R.~B. Laughlin, Phys. Rev. Lett. {\bf 50},  1395  (1983).

\bibitem{halperin84prl1583}
B.~I. Halperin, Phys. Rev. Lett. {\bf 52},  1583  (1984), \emph{ibid.}
  \textbf{52}, E2390 (1984).

\bibitem{arovas-84prl722}
D. Arovas, J.~R. Schrieffer, and F. Wilczek, Phys. Rev. Lett. {\bf 53},  722
  (1984).

\bibitem{Wilczek90}
F. Wilczek, {\em Fractional statistics and anyon superconductivity} (World
  Scientific, Singapore, 1990).

\bibitem{goldman-95s1010}
V.~J. Goldman and B. Su, Science {\bf 267},  1010  (1995).

\bibitem{depicciotto-97n162}
R. de~Picciotto, M. Reznikov, M. Heiblum, V. Umansky, G. Bunin, and D. Mahalu,
  Nature {\bf 389},  162  (1997).

\bibitem{reznikov-99n238}
M. Reznikov, R. de~Picciotto, T.~G. Griffiths, M. Heiblum, and V. Umansky,
  Nature {\bf 399},  238  (1999).

\bibitem{camino-05prl246802}
F.~E. Camino, W. Zhou, and V.~J. Goldman, Phys. Rev. Lett. {\bf 95},  246802
  (2005).

\bibitem{feldman-07prb085333}
D.~E. Feldman, Y. Gefen, A. Kitaev, K.~T. Law, and A. Stern, Phys. Rev. B {\bf
  76},  085333  (2007).

\bibitem{haldane91prl937}
F.~D.~M. Haldane, Phys. Rev. Lett. {\bf 67},  937  (1991).

\bibitem{greiter09prb064409}
M. Greiter, Phys. Rev. B {\bf 79},  064409  (2009).

\bibitem{haldane88prl635}
F.~D.~M. Haldane, Phys. Rev. Lett. {\bf 60},  635  (1988).

\bibitem{shastry88prl639}
B.~S. Shastry, Phys. Rev. Lett. {\bf 60},  639  (1988).

\bibitem{haldane91prl1529}
F.~D.~M. Haldane, Phys. Rev. Lett. {\bf 66},  1529  (1991).

\bibitem{haldane-92prl2021}
F.~D.~M. Haldane, Z.~N.~C. Ha, J.~C. Talstra, D. Bernard, and V. Pasquier,
  Phys. Rev. Lett. {\bf 69},  2021  (1992).

\bibitem{tennant-93prl4003}
D.~A. Tennant, T.~G. Perring, R.~A. Cowley, and S.~E. Nagler, Phys. Rev. Lett.
  {\bf 70},  4003  (1993).

\bibitem{coldea-97prl151}
R. Coldea, D.~A. Tennant, R.~A. Cowley, D.~F. McMorrow, B. Dorner, and Z.
  Tylczynski, Phys. Rev. Lett. {\bf 79},  151  (1997).

\bibitem{kim-06np397}
B.~J. Kim {\it et~al.}, Nature Physics {\bf 2},    (2006).

\bibitem{nayak-01prb064422}
C. Nayak and K. Shtengel, Phys. Rev. B {\bf 64},  064422  (2001).

\bibitem{misguich-02prl137202}
G. Misguich, D. Serban, and V. Pasquier, Phys. Rev. Lett. {\bf 89},  137202
  (2002).

\bibitem{moessner-01prl1881}
R. Moessner and S.~L. Sondhi, Phys. Rev. Lett. {\bf 86},  1881  (2001).

\bibitem{anderson87s1197}
P.~W. Anderson, Science {\bf 235},  1197  (1987).

\bibitem{kivelson-87prb8865}
S.~A. Kivelson, D.~S. Rokhsar, and J.~P. Sethna, Phys. Rev. B {\bf 35},  8865
  (1987).

\bibitem{Kitaev02ap2}
A.~Y. Kitaev, Ann. Phys. {\bf 303},  2  (2002).

\bibitem{kalmeyer-87prl2095}
V. Kalmeyer and R.~B. Laughlin, Phys. Rev. Lett. {\bf 59},  2095  (1987).

\bibitem{kalmeyer-89prb11879}
V. Kalmeyer and R.~B. Laughlin, Phys. Rev. B {\bf 39},  11879  (1989).

\bibitem{wen-89prb7387}
X.~G. Wen, Phys. Rev. B {\bf 40},  7387  (1989).

\bibitem{wen-89prb11413}
X.~G. Wen, F. Wilczek, and A. Zee, Phys. Rev. B {\bf 39},  11413  (1989).

\bibitem{yao-07prl247203}
H. Yao and S.~A. Kivelson, Phys. Rev. Lett. {\bf 99},  247203  (2007).

\bibitem{Kitaev06ap2}
A. Kitaev, Ann.~of Phys. {\bf 321},  2  (2006).

\bibitem{greiter-09prl207203}
M. Greiter and R. Thomale, Phys. Rev. Lett. {\bf 102},  207203  (2009).

\bibitem{Read-99prb8084}
N. Read and E. Rezayi, Phys. Rev. B {\bf 59},  8084  (1999).

\bibitem{schroeter-07prl097202}
D.~F. Schroeter, E. Kapit, R. Thomale, and M. Greiter, Phys. Rev. Lett. {\bf
  99},  097202  (2007).

\bibitem{haldane-85prb2529}
F.~D.~M. Haldane and E.~H. Rezayi, Phys. Rev. B {\bf 31},  2529  (1985).

\bibitem{AbramowitzStegun65}
{\em Handbook of Mathematical Functions}, edited by M. Abramowitz and I.~A.
  Stegun (Dover, New York, 1965).

\bibitem{laughlin89ap163}
R.~B. Laughlin, Ann. Phys. {\bf 191},  163  (1989).

\bibitem{perelomov71tmp156}
A.~M. Perelomov, Theoret. Math. Phys. {\bf 6},  156  (1971).

\bibitem{Whittaker-58}
E.~T. Whittaker and G.~N. Watson, {\em A course of Modern Analysis} (Cambridge
  University Press, Cambridge, 1958 (5th edition)).

\bibitem{Mumford83}
D. Mumford, {\em Tata Lectures on Theta} (Birkh\"auser, Basel, 1983), Vol.~I
  and II.

\end{thebibliography}

\end{document}